\documentclass[12pt]{article}
\usepackage{amsmath,amssymb}
 \usepackage{graphicx}

\usepackage{graphicx}
\usepackage{wrapfig}

\usepackage{hyperref}
\usepackage{cite}

\def\beq{\begin{eqnarray}}
\def\eeq{\end{eqnarray}}

\newcommand{\Tr}{\,\mathrm{Tr}\,}            

\newcommand{\be}{\begin{equation}}
\newcommand{\ee}{\end{equation}}
\newcommand{\bea}{\begin{eqnarray}}
\newcommand{\eea}{\end{eqnarray}}
\newcommand{\bg}{\begin{gather}}

\newcommand{\bseq}{\begin{subequations}}
\newcommand{\eseq}{\end{subequations}}

\newcommand{\sm}[1]{{\scriptscriptstyle \rm #1}}

\def\be{\begin{eqnarray}}
\def\ee{\end{eqnarray}}
\def\lb{\label}

\begin{document}
\title{\textbf{Renormalization group equations \\  and  the  recurrence    pole relations  \\  in pure  quantum gravity
}}

\author{ \textbf{Sergey N. Solodukhin}} 

\date{}
\maketitle
\begin{center}
\hspace{-0mm}
    \emph{  Institut Denis Poisson UMR 7013,
  Universit\'e de Tours,}\\
  \emph{Parc de Grandmont, 37200 Tours, France} \\
\end{center}



\maketitle \thispagestyle{empty}
\vspace*{.5cm}

\begin{abstract}
\noindent {In the framework of dimensional regularization,  we propose a  generalization of  the renormalization group equations in the case of the perturbative quantum gravity that
 involves renormalization of the metric and of the higher  order Riemann curvature couplings. The case of zero cosmological constant is considered.
Solving the renormalization group (RG) equations we compute the respective beta functions and derive the recurrence relations, valid at any order in the  Newton constant,   that relate the higher pole  terms $1/(d-4)^n$
to a single pole $1/(d-4)$  in the quantum  effective action. Using the recurrence relations we find the exact form for the higher pole counter-terms that appear in 2, 3 and  4 loops and 
we make certain statements about the general structure of the higher pole counter-terms in any loop. 
We  show that the complete set of the UV divergent terms can be  consistently (at any order in the Newton constant) hidden in the bare gravitational action, that includes the terms of higher order in the Riemann tensor, provided the metric and the higher curvature couplings are renormalized according to the RG equations.}

\end{abstract}

\vskip 2 cm
\noindent
\rule{7.7 cm}{.5 pt}\\ sergey.solodukhin@lmpt.univ-tours.fr
\noindent 
\noindent

\pagebreak

\newpage

\section{ Introduction}
\setcounter{equation}0
\bigskip

The ultra-violet (UV) divergences in a quantum field theory are handled by adding the respective counter-terms. The structure of the UV  divergences is
most conveniently analyzed in the framework of the dimensional regularization that preserves all symmetries present in the quantum system. The divergences then appear as
a set of poles when the space-time dimension $d\rightarrow 4$. The respective terms in the quantum effective action are not independent: the higher pole terms $1/(d-4)^n\, , \, n>1$  are completely determined by the terms in a single pole $1/(d-4)$. These relations, known in the literature as the pole equations, come out as a consequence of the Renormalization Group 
equations as was demonstrated by 't Hooft \cite{tHooft:1973mfk}.  
Earlier the pole equations were derived and used, quite effectively,  in the case of 2d  sigma-models in which one  has to renormalize the metric in the target space \cite{Friedan:1980jm}, \cite{AlvarezGaume:1981hn}.  

In the case of  the perturbative quantum gravity the computation of the loop diagrams is notoriously difficult.  The one- and two-loop results are the only available in the literature \cite{'tHooft:1974bx}, \cite{Goroff:1985th}, \cite{vandeVen:1991gw} although the possible structure of the counter-terms that may in principle appear in the higher loops  can be
 analyzed by means of the covariance principle, at least at some lower orders in the curvature. Their number  is rapidly increasing with number of loops and in all orders in the Newton constant one  deals with an infinite number of possible structures.

It was suggested in  \cite{Kazakov:1987jp}  (see also \cite{Kazakov:1987ej} and for some later  developments  \cite{KK})  that the Renormalization Group methods should be  equally applicable  to the non-renormalizable theories.
This program if successful would make the non-renormalizable theories   to look quite similar  to the renormalizable ones. 
 The available  in the literature prescriptions, however, are rather implicit.
 We do not directly  rely here on this previous work although it played an important inspirational role for our study.

In the present paper we develop a systematic  approach to perturbative  quantum gravity that uses the renormalization group equations. 
The primary goal of the paper is to derive the recurrence pole relations in the case of gravity. 
Our prescriptions are precise and unambiguous.
It should be noted that over the  last several  decades there have been 
suggested a number of approaches to quantum gravity that refer to  certain versions of the  Renormalization Group, in most cases of the Wilsonian type. In order to 
avoid any possible confusion we would like to 
stress  from the very beginning that none of these approaches will be  used here. The closest analogue of the approach that we develop in the present paper is that of the renormalization group  equations of 't Hooft \cite{tHooft:1973mfk}. The key important point in our construction, that, to the best of our knowledge, was missing in the earlier approaches to quantum gravity, is the necessity to consider a renormalization of the metric, much in the same way as in a renormalizable QFT
one introduces a renormalization of the quantum fields. The peculiarity of this renormalization procedure for the metric is that  it is not multiplicative but of a rather general, although still local,  type. 

The other important remark is that throughout the paper only the case of pure gravity with zero cosmological constant will be considered.
The case of non-zero cosmological constant will be treated in a subsequent work.

The paper is organized as follows. In section 2 we briefly review the method of 't Hooft  in the case of a renormalizable quantum field theory (QFT).  In section 3 we derive the renormalization group equations in the case of pure quantum gravity. In section 3.1 we consider the renormalization of the metric  and in section 3.2 the renormalization of the quantum effective action. In section 4 we solve some of the RG equations and 
determine the exact form of the higher loop metric beta function and the beta functions for the higher order curvature coupling constants. In section 5 we derive the pole recurrence relations and 
solve these relations to determine the exact form of the higher pole counter-terms in 2, 3 and 4 loops.
In section 6 we focus on the General Relativity (GR) counter-terms and make some  statements on their general form. In section 7 we demonstrate that, similarly to the renormalizable theories,  the complete set of the UV divergences can be hidden in the bare  gravitational action
provided the bare metric  and the bare higher curvature couplings are  expressed in terms of the renormalized quantities. 
We conclude in section 8.

 \section{RG equations in renormalizable QFT}
 \setcounter{equation}0
\bigskip
 Before we start our analysis  of quantum gravity we briefly review    the derivation of the renormalization group 
 equations in 't Hooft's method  in the case of a  4d renormalizable theory  \cite{tHooft:1973mfk}, see also \cite{AlvarezGaume:1981hn},   \cite{Kazakov:1987jp} for a similar review.
 Consider a dimensionless coupling constant $\lambda$. In  $d=4-\epsilon$  space-time dimensions the bare coupling $\lambda_B$ has dimension $[\mu^{\epsilon}]$, where $\mu$ is mass scale, as for instance in the case of $\phi^4$ theory. In dimensional regularization one develops a series of counter-terms of the classical action such that the bare coupling constant is expressed  as a function  of a dimensionless renormalized 
 coupling $\lambda_R$,
 \be
 \lambda_B=\mu^{\epsilon}(\lambda_R+\sum_{k=1}\frac{a_k(\lambda_R)}{\epsilon^k})\, .
 \lb{r-1}
 \ee
 The renormalized coupling $\lambda_R$ is a function of scale $\mu$ such that an equation
 \be
 \mu\partial_\mu \lambda_R=-\epsilon\lambda_R+\beta(\lambda_R)\, 
 \lb{r-2}
 \ee
 holds.  The bare coupling is supposed to be independent of $\mu$ so that $\mu\partial_\mu \lambda_B=0$. Differentiating equation (\ref{r-1}) with respect to $\mu$ one obtains the following equation
 \be
 \epsilon\sum_{k=1}\frac{a_k}{\epsilon^k}+\beta(\lambda_R)+(-\epsilon \lambda_R+\beta(\lambda_R))\sum_{k=1}\frac{a'_k(\lambda_R)}{\epsilon^k}=0\, ,
 \lb{r-3}
 \ee
 where $a'_k(\lambda_R)\equiv \partial_{\lambda_R}a_k(\lambda_R)$ and the terms linear in $\epsilon$  cancel out. 
 
 The constant  term $\epsilon^0$ in (\ref{r-3}) gives us equation that allows to express the beta function in terms of a single pole $a_1$,
 \be
 \beta(\lambda_R)=a_1(\lambda_R)-\lambda_R a'_1(\lambda_R)\, .
 \lb{r-4}
 \ee
 The vanishing condition for a pole  $1/\epsilon^k\, , \ k\geq 1$ in  eq. (\ref{r-3}) produces a recurrence relation
 \be
 a_{k+1}(\lambda_R)-\lambda_R a'_{k+1}(\lambda_R)=\beta(\lambda_R)a'_k(\lambda_R)\, , \ \ k\geq 1\, 
 \lb{r-5}
 \ee
 This relation together with the beta function (\ref{r-4}) form a set of recurrence relations that uniquely determine the higher pole residues  $a_k$ in (\ref{r-1}) provided a single pole  residue $a_1$ is given.
 Similar equations can be written for the renormalization of masses and the quantum fields  \cite{tHooft:1973mfk}. The renormalization of fields   should not be necessarily multiplicative in general. 
 Besides other things, the pole equations play the role of the  consistency conditions to be satisfied in the higher loop calculations. Below we generalize these equations in the case of pure quantum gravity
 without a cosmological constant.

\section{ Renormalization group equations in pure quantum gravity}
\setcounter{equation}0
\bigskip
Our starting point is the theory of gravitational field described by the action
\be
L_{gr}=-\frac{1}{G_N}\int R\sqrt{g}d^4x=\frac{L_0}{G_N}\, ,
\lb{1}
\ee
where $G$ is the Newton constant (notice that we absorb the usual factor $16\pi$ in the definition of $G_N$). This action does not include the cosmological constant
that is assumed to vanish. As we have mentioned this above the case of non-vanishing cosmological constant deserves a separate study and will be reported later.
Thus,  there is only one dimensionful parameter, the Newton constant $G_N$.

\subsection{Renormalization of the metric}

In the dimensional regularization one considers the  space-time dimension $d$ slightly different from $4$. In doing so the otherwise dimensionless quantities
acquire some dimension that can be compensated by introducing a new scale $\mu$. In what follows we prefer to keep the dimensionality of the
Newton constant to be the same as in $d=4$, i.e. $[G_N]=2$. Instead, the bare metric gets some dimensionality $[g_{B,ij}]=-(d-4)$ and it is the only quantity present in the classical action (\ref{1}) that has to be renormalized.
This renormalization is in fact a field renormalization that takes  a general local form, 
\be
g_{B,ij}=\mu^{-\epsilon} ( g_{R,ij}+\sum_{k=1} \epsilon^{-k}\, h_{k, ij}(g_R))\, , 
\lb{2}
\ee
where $\epsilon=4-d$ and $g_{R,ij}$ is a dimensionless renormalized metric, $h_{k,ij}(g_R)$ are local covariant functions of the renormalized metric. 
One has that
\be
\mu\partial_\mu g_{R,ij}=\epsilon g_{R,ij}+\beta_{ij}(g_{R})\, ,
\lb{3}
\ee
where the beta function $\beta_{ij}(g_{R})$ is a local function of $g_{R}$. 
 Eqs.(\ref{2})-(\ref{3}) are quite similar to the renormalization of the target metric  in the $d=2$ sigma-models \cite{Friedan:1980jm}, \cite{AlvarezGaume:1981hn}.
 We should, however, stress the obvious differences: the target metric in a sigma model represents an infinite set of couplings while here we deal with a field renormalization. 

The bare metric $g_{B,ij}$ is independent of the scale $\mu$ so that differentiating the both sides of equation (\ref{2}) and using (\ref{3}) we arrive at the equation
\be
-\epsilon \sum_{k=1}\frac{h_{k}}{\epsilon^k}+\beta(g_R)+\sum_{k=1}\frac{1}{\epsilon^k}h'_{k}(g_R)\times (\epsilon g_{R}+\beta(g_R))=0\, ,
\lb{4}
\ee
where we skip the space-time indices and the terms linear in $\epsilon$   already canceled out.  Let us explain our notations used in the above expression:
for a tensor $h_{ij}(g)$, a local covariant function of the metric, and a tensor $f_{ij}(x)$ we define
\be
(h'\times f)_{ij}\equiv \frac{d}{dt}h_{ij}(g+t\, f)|_{t=0}\, .
\lb{4-11}
\ee

Matching the coefficients for a constant term in (\ref{4}) one finds the expression for the beta function in terms of a single pole term,
\be
\beta(g_R)=h_{1}(g_R)-h'_{1}(g_R)\times g_{R}\, .
\lb{5}
\ee
For the coefficients at a higher pole $1/\epsilon^k\, , \, k\geq 1$ one finds
\be
h_{k+1}-h'_{k+1}\times g_R=h'_{k}\times \beta(g_R)\, , \, k\geq 1\, .
\lb{6}
\ee
This is a recurrence relation for the higher pole residues in terms of  the lower pole residues. As always in the case of the RG equations, the complete information  about the renormalization is contained
in a single pole $h_{1}$. In the perturbative expansion with respect to the Newton constant each term can be expressed as a power series in $G_N$, 
\be
h_{k}=\sum_{l=k}G_N^l \, h_{k,l}\,  \, , \, \, \, \beta_{ij}=\sum_{l=1}G_N^l\beta_{l, ij}\, ,
\lb{7}
\ee
where $h_{k,l}$ is a local polynomial in curvature of degree $l$ (we count two covariant derivatives acting on a curvature  to have same degree $1$ as the curvature itself).
Notice that $h'_g (g)g$ is a transformation of $h(g)$ under the rescaling of the metric, $g\rightarrow \lambda g$. So that one finds that $h'_{k,l}(g)\times g=(1-l)h_{k,l}$.
Applying this to equations (\ref{5}) and (\ref{6}) one finds for the terms in the expansion (\ref{7}),
\be
\beta_l=l\, h_{1,l}  \, , \  \ \ h_{k+1,l}=\frac{1}{l}\sum_{p=1}^{l-1}h'_{k,l-p} \times \beta_p=\frac{1}{l}\sum_{p=1}^{l-1}p\, h'_{k,l-p} \times h_{1,p}\, , \ \ k\geq 1\, .
\lb{8}
\ee
This is first set of the  RG equations that we will deal with. The other one arises when the renormalization of the effective gravitational action is considered.
It will be discussed in the next section. We finish this section by saying that the solution to the  RG equations for the metric that we have just  derived is completely determined
by either specifying the single pole terms $\{h_{1,l}\}$ or the beta function terms $\{\beta_{l}\}$ in the decomposition (\ref{7}). In the two-dimensional sigma-models the primary element is a single pole term
that determines the beta function. In the present case it is rather the beta functions $\{\beta_l\}$ that we have to specify first and then determine all terms in the expansion (\ref{2}), (\ref{7}).
However, in order to determine the beta function for the metric we have to look at the renormalization of the action. The metric beta function is then completely specified
(up to the gauge coordinate transformations that we will discuss)  by a single pole in the effective action. These issues are discussed in the next section.

\subsection{Renormalization of the action}

The renormalized gravitational action with all counter-terms added is a function of the renormalized metric $g_{R}$. The counter-terms are divided on two classes:
those that vanish on-shell, we call them $L_k$, and those that do not, we call them $V_k$. $V_k$ are invariants constructed from the Riemann  tensor and its covariant derivatives
that remain non-trivial provided the Bianchi identities are used. It is not a goal of the present paper to give a classification of all possible non-equivalent curvature invariants of a given order.
We, however, assume that this classification can be done and, possibly,  already exists in the mathematical literature although we are not aware of any relevant publications.
In the classical part of the action we have to add terms that are due to the Riemann tensor only, we call them $W$,
with the appropriate coupling constants that have to be renormalized in order to  absorb the UV divergences due to $V_k$. It is convenient to choose a basis of integral invariants
of degree $(l+1)$ constructed from the Riemann tensor and its covariant derivatives, $\{P_l \, , \, l\geq 2\} $, and expand $W$ and $V_k$ with respect to this basis
\be
W=\sum_{l=2}  W_l, , \ \ \  V_k=\sum_{l\geq k}G_N^{l-1}V_{k,l}\, , \ \ \ W_l=G_N^{l-1}\lambda_l P_l \, , \ \ \ \  V_{k,l}=v_{k,l}P_l\, ,
\lb{9}
\ee
where $\{\lambda_l\}$ is a set of dimensionless coupling constants and, since for each $l$ there will be a certain number of independent invariants,  both $\lambda_l$ and $v_{k,l}$ are supposed to have  an extra index  
to enumerate the different invariants of same degree $l+1$.  In $d=4$ the invariant quadratic in the Riemann tensor can be expressed in terms of invariants quadratic in the Ricci tensor and Ricci scalar 
and the Euler topological invariant. Therefore,  in first order $(l=1)$ the UV divergent terms vanish on-shell and one has that $v_{1,1}=0$. In the second order (two loops),  $l=2$,  there is only one independent invariant
that  can be constructed from the Riemann tensor, $P_2=\int d^4 x \sqrt{g}R_{ab}^{\ \ cd}R_{cdmn}R^{mnab}$. The  numerical value of $v_{1,2}$ was first 
computed  by Goroff and Sagnotti  \cite{Goroff:1985th} and later by van de Venn \cite{vandeVen:1991gw}.

Similarly to (\ref{9}) we expand the counter-terms $L_k$ in powers of the Newton constant $G$,
\be
L_k=\sum_{l\geq k}G_N^{l-1}L_{k,l}\, ,
\lb{9-1}
\ee
where $L_{k,l}$ contains the curvature invariants of degree $l+1$. The terms $L_{k,l}$ vanish on-shell and, hence, have to contain at least one power
of the Ricci tensor or Ricci scalar.

\subsubsection{Higher order Riemann curvature  terms in gravitational action}

\begin{figure}[!t]
\centering
\includegraphics[width=100mm]{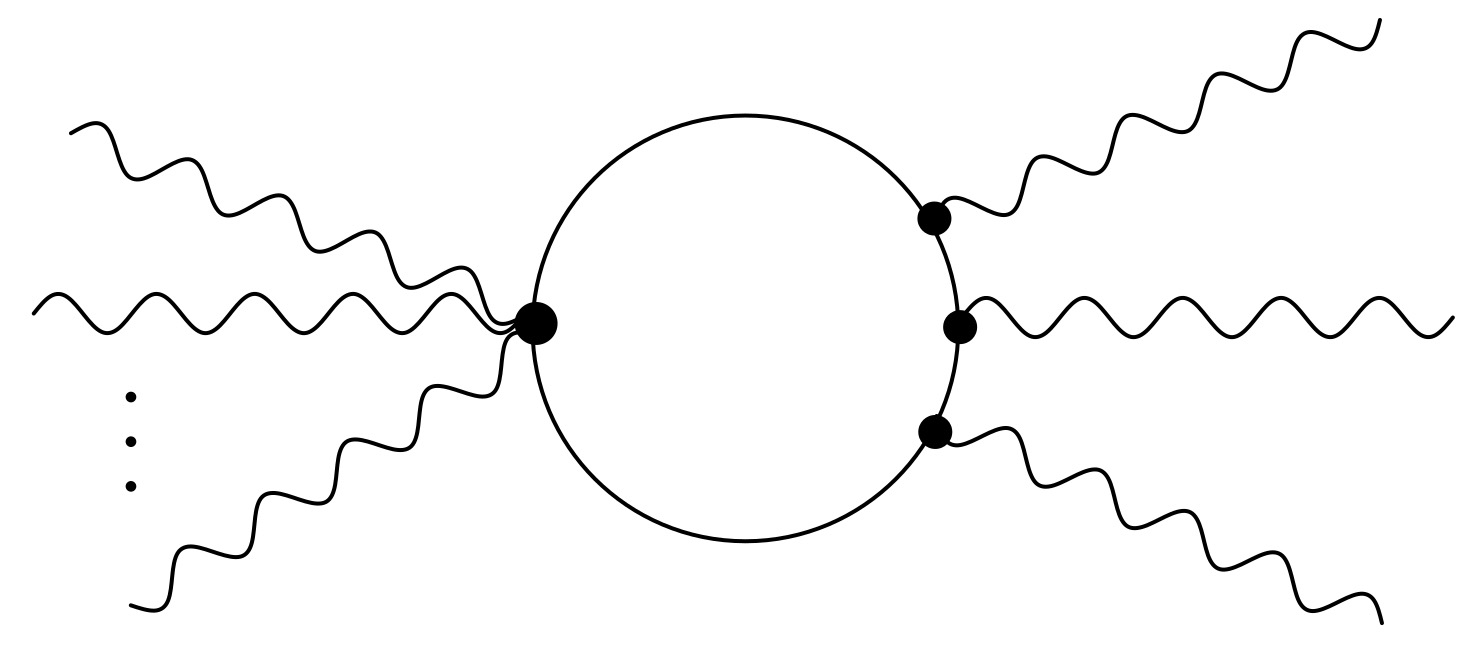}
\caption{One-loop graviton diagram with three GR cubic vertices  and one   $l$-point vertex  due to a higher curvature  term.}
\label{Feynman}
\end{figure}

It should be noted that by adding $W$ to the action $L_0$ one  does not change the graviton propagator.  Indeed, expanding metric over Minkowski spacetime, $g_{ij}=\eta_{ij}+\sqrt{G}\phi_{ij}$, where $\phi_{ij}$ is a perturbation, one finds that the terms
\be
W_l\sim \Lambda_l(\sqrt{G})^{l+1} \int d^4x(\partial\partial\phi)^{l+1}\, , \ \ l\geq 2
\lb{9-11}
\ee
where $\Lambda_l=G^{l-1}\lambda_l$, start with a term of $(l+1)$-th order in perturbation and, since $l\geq 2$,   do not contribute to the graviton propagator. These new terms lead to an additional  $(l+1)$-point vertex with a new coupling in the graviton Feynman diagrams.  Each leg in the vertex has two spacetime derivatives.
The corresponding coupling constant is $\Lambda_l (\sqrt{G})^{l+1}$.  For $l=2$ this is a $3$-point vertex with two space-time derivatives at each leg. We remind that  a (usual) $3$-point vertex in General Relativity has effectively  a coupling constant $\sqrt{G}$ and at maximum it has one leg with two derivatives.
Effectively, the  expansion will go with respect to all available coupling constants: $G\, , \ \Lambda_2\, , \ \Lambda_3\, , \ \dots$.
We do not give here a detailed  analysis of the corresponding  UV divergences.
It is, however, instructive to observe certain rules by looking at some simple examples and by using the dimensionality arguments.
 We  first remark that by the dimensionality the UV divergent term that is linear in $\Lambda_p$ and that contains $(l+1)$-th power of the  curvature has the following form,  
  \be
  \sum_{p=2}^{l-1} \Lambda_p G^{l-p} \int d^4 x {\cal R}^{l+1}\, ,
  \lb{cc}
  \ee
  $\cal R$ stands here for any curvature, the Ricci tensor or Riemann tensor.
 A one-loop diagram that may produce a UV divergent term  of this type for $p=l-1$ is shown in Fig.\ref{Feynman}. 
It contains one $\Lambda_p$ vertex and three GR 3-point vertices. 
 For other values of $p$ the diagram would include a graviton vertex of $V_p$ and $r=(l-1)-p$ internal GR graviton lines.
 It is clear that in the renormalization of the coupling $\Lambda_l$ there may appear only the couplings $\Lambda_p$ with $p\leq l-1$. So that the upper limit in the sum in (\ref{cc}). 
 Same restriction comes from the condition that the number of the internal graviton  lines $r\geq 0$.
A similar analysis shows that a Feynman diagram with $\Lambda_{p_1}\, , \ \Lambda_{p_2}\, , \dots\, , \ \Lambda_{p_n}$ vertices, $r$ internal GR lines and $m$ GR vertices such that
$p_1+\dots+p_n-n+m=l+1$ produces a UV divergent term of $(l+1)$-th order in curvature,
\be
\Lambda_{p_1}\dots \Lambda_{p_n}G^{(n+r)}\int d^4x{\cal R}^{l+1}\, , \ \  p_1+\dots+p_n=(l-1)-r\, .
\lb{x}
\ee
The counter-term has to have dimension zero, so that one gets a condition for $p_i$ as above.
Since $r\geq 0$ one has a condition on the values of $p_i$: 
\be
p_1+\dots+p_n\leq l-1\, .
\lb{z}
\ee
This discussion  is not a rigorous proof of validity of this bound in general. So that its status is conjectural.   Although we will not need it in the most of our consideration below  this
bound will help to avoid certain ambiguities in the beta function equations that will be discussed in section 7.2.
Thus,   the lowest parameter $v_{1,2}$ (or the counter-term $V_{1,2}$) is independent of any $\lambda$. On the other hand, the higher order  counter-terms,  $V_{k,l}$ and $L_{k,l}\, , \, l\geq 3$ can be  polynomial functions of $\lambda_p\, , p\leq l-1$
provided the condition (\ref{z}) is satisfied. We see that, when the higher curvature terms are present,  the lower loops may give contributions to the curvature terms that  appear at the GR loop order $l$.
For convenience, we will still refer to $l$ as a loop order.

\subsubsection{RG equations for  $\lambda$ couplings }

In $d$ space-time dimensions the bare coupling $\lambda^B_l$ has dimension $\mu^{l(d-4)}$.
Expressing the bare couplings in terms of dimensionless renormalized quantities $\lambda_l^R$ one has
\be
\lambda^B_l=\mu^{-l\epsilon}\left( \lambda^R_l+\sum_{k=1}\epsilon^{-k}a_{k,l}(\lambda^R)           \right)\, , \ \  l\geq 2
\lb{9-2}
\ee
The renormalized couplings $\lambda^R_l$ satisfy equation
\be
\mu\partial_\mu\lambda^R_l= \epsilon\, l\,\lambda^R_l+\hat{\beta}_l \, ,
\lb{10}
\ee
where $\hat{\beta}_l$ is  the beta function for  coupling $\lambda_l$. The renormalization group equations for the couplings $\lambda_l\, , \  l\geq 2$ read
\be
&&\hat{\beta}_l=la_{1,l}-\sum_{p=2}^{l-1} p\lambda_p\partial_{\lambda_p}a_{1,l}\, , \nonumber \\
&&\left(1-\frac{1}{l}\sum_{p=2}^{l-1}p\lambda_p\partial_{\lambda_p}\right)a_{k+1,l}=\sum_{p=2}^{l-1}\hat{\beta}_p\partial_{\lambda_p}a_{k,l}\, .
\lb{10-1}
\ee
Here we take into account that  in the renormalization of a coupling $\lambda_l$ there may be involved only the couplings $\lambda_p$ with $p\leq l-1$.

\subsubsection{Modified RG equations for metric}

In general, the bare metric can be a function of the renormalized couplings $\lambda^R$. So that the terms $h_{(k)}$ in the expansion (\ref{2}) are, in general, functions of both the renormalized metric $g_{R}$ and the renormalized couplings $\{\lambda^R_l\}$. The condition $\mu\partial_\mu g_{B}=0$ then leads to the modified RG equations,
\be
&&\beta_1=h_{1,1}\, , \ \  \beta_l=l\, h_{1,l}-\sum_{p=2}^{l-1}p\lambda_p\partial_{\lambda_p}h_{1,l}\, , \ l\geq 2 \\
&&\left(1-\frac{1}{l}\sum_{p=2}^{l-1}p\lambda_p\partial_{\lambda_p}\right)h_{k+1,l}=\frac{1}{l}\sum_{p=1}^{l-1}h'_{k,l-p}\times \beta_p+\frac{1}{l}\sum_{p=2}^{l-1}\hat{\beta}_p\partial_{\lambda_p}h_{k,l}\, , \ k\leq l-1\nonumber
\lb{10-3}
\ee
These equations present a  modification of the metric RG equations considered in section 3.1.

\subsubsection{RG equations for quantum action}

In the renormalizable theories the renormalization procedure goes in few steps. The quantum action, which is a sum of the classical action and the counter-terms, is supposed to be a function of the renormalized fields and  the renormalized couplings and masses.  It does not depend on the scale $\mu$. This condition imposes  certain equations on the residues of the poles $1/(d-4)^k$ in the quantum action.  Then, in the renormalizable theories the quantum action 
takes the form of the classical (bare) action provided it is expressed in terms of the bare fields, couplings and masses. The latter, on the other hand, are functions of the renormalized quantities.
These steps can be repeated in a non-renormalizable theory. The RG equations for the quantum action in a rather general theory were previously considered in \cite{Kazakov:1987jp}. However, the construction in \cite{Kazakov:1987jp}
was not accompanied, in a coherent way,  by a suitable renormalization of the fields and couplings. Their equations are different from those considered in the present paper.

The quantum gravitational action is a function of all renormalized quantities, the metric and the higher curvature couplings, and  it  takes a  general form 
\be
L_Q(g_R(\mu),\lambda^R_l(\mu))=\mu^{-\epsilon}\left( \frac{1}{G}L_0(g_R)+W(g_R, \lambda_R)+\sum_{k=1}\frac{1}{\epsilon^k}(L_k(g_R,\lambda_R)+V_k(g_R,\lambda_R))\right)\, .
\lb{12}
\ee
The power of $\mu$ is uniquely determined by the requirement that the action to have dimension zero.
The quantum action does not depend on the scale $\mu$, the differential equation $\mu\partial_\mu L_Q=0$ leads to  equation
\be
&&-\epsilon ( \frac{1}{G_N}L_0+\sum_{k=1}\frac{1}{\epsilon^k}(L_k+V_k))+\frac{1}{G_N}L'_0\cdot (\epsilon g+\beta)+\sum_{k=1}\frac{1}{\epsilon^k}(L'_k+V'_k)\cdot (\epsilon g+\beta)\nonumber \\
&&+\sum_{l=2}G_N^{l-1}[-\epsilon\lambda_l P_l+\lambda_l P'_l\cdot (\epsilon g+\beta)+P_l (\epsilon\, l\, \lambda_l+\hat{\beta}_l)]\nonumber \\
&&+\sum_{k=1}\frac{1}{\epsilon^k} \sum_{p=2}\partial_{\lambda_p}
(L_k+V_k)(\epsilon p \lambda_p+\hat{\beta}_p)
=0\, ,
\lb{13}
\ee
where  in order to simplify the expression we drop the subscript $R$ in the renormalized metric $g_R$ and in the renormalized higher curvature couplings $\lambda^R_l$, the second line in (\ref{13}) is due to differentiation of $W$.
 In order to simplify further the formulas, throughout the paper, we  will maximally use the index-free notations. 
 Let us explain our notations used in the above expression. Each term in the action (\ref{12})
is a  functional of the renormalized metric $g_R$.  $L'$ stands for the metric variation so that it is a local tensor $(L')^{ij}=\frac{\delta L}{\delta g_{ij}(x)}$.  In particular, $(L'_0)^{ij}=(R^{ij}-\frac{1}{2}g^{ij}R)=G^{ij}$ is the Einstein tensor.  Then we define
\be
 L' \cdot \beta \equiv \frac{d}{dt}L[g+t\beta]|_{t=0}=\int d^4x\sqrt{g}\, (L')^{ij}\beta_{ij}\, . 
 \lb{0}
 \ee
Each term in the action (\ref{12}) is an integral  over spacetime.  Therefore, we will systematically neglect any total derivatives that may appear under the integral.

We notice that the equation (\ref{13})  is invariant under a redefinition of the metric beta function,
\be
\beta_{ij}\rightarrow \beta_{ij}+\nabla_i\xi_j+\nabla_j\xi_i\, .
\lb{x}
\ee
This is a usual ambiguity for the metric beta function. We stress that the metric beta function in equation (\ref{13}) is the one that we studied in  Section 3.1, see  (\ref{3}) and   (\ref{10-3}).

Consider a rescaled metric $\lambda g_{ij}$.
Then with our notations one has that  for any functional of the metric $\frac{dW}{d\lambda}|_{\lambda=1}=W'\cdot g$. In particular, this gives us a relation
\be
T'_l\cdot g =-(l-1) T_{l}\, 
\lb{14}
\ee
for any curvature polynomial of degree $l+1$ (any two covariant derivatives are counted as one curvature degree). In particular, this relation holds for $T_l=P_l$, polynomials of the Riemann tensor and its covariant derivatives. 
In the first line of (\ref{13}) the linear in $\epsilon$ terms cancel   due to relation $-L_0+L_0'\cdot g=0$ that is an extension of (\ref{14}) for $l=0$.

\section{ Beta functions}
\setcounter{equation}0
\bigskip
The vanishing of a  constant, $\epsilon^0$, term in (\ref{13}) will give us the relations to determine   the beta functions $\beta_{ij}$ and $\hat{\beta}_l$.
One has that
\be
L_1+V_1-\frac{1}{G}L_0'\cdot \beta-(L'_1+V'_1)\cdot g-\sum_{l=2}G^{l-1}(\lambda_lP'_l\cdot\beta+P_l\, \hat{\beta}_l)-\sum_{p=2}p\lambda_p\partial_{\lambda_p}( L_1+V_1)=0
\lb{15}
\ee
In order to proceed further we use the expansion of the counter-terms in series with respect to the Newton constant $G$, (\ref{9}), (\ref{9-1}). One finds that
\be
l L_{1,l}-L'_0\cdot\beta_l-P_l\, (\hat{\beta}_l-lv_{1,l})-\sum_{m=2}^{l-1}\lambda_m P'_m\cdot \beta_{l-m}-\sum_{p=2}^{l-1}p\lambda_p\partial_{\lambda_p}(L_{1,l}+V_{1,l})=0\, ,
\lb{16}
\ee
where $l\geq 1$ and we used (\ref{14}). Notice that the 3rd term in (\ref{16}) is non-trivial for $l\geq 2$ and the 4th  and 5th terms are non-zero for $l\geq 3$. 

We  will use the following general representation for terms $L_{1,l}$
\be
L_{1,l}=L'_0\cdot X^{(l)}=\int d^4x\sqrt{g} G^{ij}X^{(l)}_{ij}\, ,
\lb{16-1}
\ee
where $X^{(l)}_{ij}$ is a curvature polynomial of degree $l$ (any two covariant derivatives are counted as one curvature degree).
We note that, provided that $L_{1,l}$ is given,  the term $X^{(l)}_{ij}$ is determined up to a redefinition 
\be
X^{(l)}_{ij}\rightarrow X^{(l)}_{ij}+\nabla_i \xi_j+\nabla_j\xi_i\, .
\lb{16-2}
\ee

\medskip

\noindent Below we consider some particular values of $l$.

\bigskip

\noindent \underline{{\it \bf $l=1$}}

\medskip

Equation (\ref{16}) in this case  reduces to only first two terms,
\be
-L_{1,1}+L'_0\cdot \beta_{1}=0\, .
\lb{17}
\ee
In one loop, the quantum effective action contains terms quadratic in the Ricci scalar and in the Ricci tensor (the square of the Riemann tensor reduces to these two invariants plus a topological Euler number that we neglect in our study).
We represent 
\be
L_{1,1}=\int d^4 x \sqrt{g}(aG_{ij}^2+bG^2)=\int d^4 x\sqrt{g}\, G^{ij}X_{ij}^{(1)}\, , \, X^{(1)}_{ij}=a\,G_{ij}+b\,g_{ij}G\, ,
\lb{18}
\ee
where $G_{ij}=R_{ij}-\frac{1}{2}g_{ij}R$, $G=g^{ij}G_{ij}=-R$. Values of $a$ and $ b$ are available in the literature and are known to depend on the gauge.
The equation for the beta function $\beta_1$  is 
\be
L'_0\cdot X^{(1)}=L'_0\cdot \beta_{1}\, .
\lb{19}
\ee
A solution of this equation  is
\be
(\beta_1)_{ij}=X^{(1)}_{ij}=a\,G_{ij}+b\,g_{ij}G\, .
\lb{20}
\ee
This solution is not unique. One can add to (\ref{20}) a term of the form $\nabla_i \xi_j+\nabla_j\xi_i$ where $\xi_i$ is  an arbitrary vector field.
This is a general ambiguity (\ref{x}) for the beta function of the metric. 
We notice that at this order the beta function (\ref{20}) vanishes on-shell ($G_{ij}=0$).

\bigskip

\noindent \underline{\bf $l=2$}

\medskip

In this order equation (\ref{16}) has more terms,
\be
-2L_{1,2}+L'_0\cdot \beta_2+P_2\, (\hat{\beta}_2-2v_{1,2})=0\, .
\lb{21}
\ee
We remind that, by definition, $L_{1,2}$ vanishes on-shell and hence can be presented in the form
\be
L_{1,2}=L'_0\cdot X^{(2)}=\int d^4 x \sqrt{g} G^{ij}X^{(2)}_{ij}\, ,
\lb{22}
\ee
where $X^{(2)}$ is quadratic in curvature, it does not necessarily vanish on-shell, in particular it may contain a term quadratic in the Riemann tensor (see below). 
On the other hand, $P_2$ is invariant cubic in the Riemann tensor. In $d=4$ there is only one such invariant,  
\be
P_2=\int d^4 x\sqrt{g}R_{ijkl}R^{klmn}R_{mn}^{\ \ \ ij}\, .
\lb{22-1}
\ee
One sees that in (\ref{21}) the first two terms vanish on-shell and the last term does not vanish. This means that 
the first two terms and the last term in the equation are independent and should vanish separately. This gives us the two loop beta functions for the metric and for the cubic coupling 
$\lambda_2$,
\be
(\beta_2)_{ij}=2X^{(2)}_{ij}\, , \  \ \ \hat{\beta}_2=2v_{1,2}\, .
\lb{23}
\ee
Tensor $X^{(2)}_{ij}$ is a local tensor quadratic in curvature or its covariant derivatives. Its general form is
\be
&&X^{(2)}_{ij}=c_0\,  g_{ij}R_{nklm}R^{nklm}+G_{kl}Y^{kl}_{ij}\, , \nonumber \\
&&Y^{kl}_{ij}=c_1 R_{i\  \  \ j}^{\ (k l)}+c_2 (g^{kl}G_{ij}+\delta^{(k}_i\delta^{l)}_j G)+c_3 G^{kl}g_{ij}+c_4g^{kl}g_{ij}G+c_5G^{(k}_i\delta^{l)}_j\, ,
\lb{24}
\ee
where we did not include the term with covariant derivatives of scalar curvature since it has the form of a gauge transform (\ref{16-2}). 
We also did not include a product of two Riemann tensors with two free indices since in $d=4$ it is expressed in terms of other curvature invariants, 
\be
R_{i}^{\ nkl}R_{jnkl}=2 R_{ikjl}R^{kl}-RR_{ij}+2R_{ik}R_{j}^{\ k}+\frac{1}{4}g_{ij}(R^2_{nklm}-4R^2_{kl}+R^2)\, .
\lb{25}
\ee
The simplest way to get this relation is to vary the $d=4$ Euler density, see for instance \cite{Gurses:2020ofy} and, for an alternative derivation in terms of the Weyl tensor, \cite{Barvinsky:1994hw}.
When derived (\ref{24}) we also took into account the fact that a certain combination of tensors  is orthogonal to the Einstein tensor $G_{kl}$,
\be
G_{kl}\tilde{Y}^{kl}_{ij}=0\, , \ \ \  \tilde{Y}^{kl}_{ij}=g^{kl}G_{ij}-\delta^{(k}_i\delta^{l)}_j G\, .
\lb{25-1}
\ee
We see that in  two loops there may appear a term (proportional to $c_0$) in the metric beta function that does not vanish on-shell.  
The term due to $c_1$ is linear in $G_{ij}$ and the other terms are quadratic in $G_{ij}$.

\bigskip

\noindent \underline{\bf $l=3$}

\medskip

In the cubic order (three loops) all terms in equation (\ref{15}) contribute,
\be
-3L_{1,3}+L'_0\cdot\beta_3+P_3\, (\hat{\beta}_3-3v_{1,3})+\lambda_2\, P'_2\cdot\beta_1 +2\lambda_2\partial_{\lambda_2}(L_{1,3}+v_{1,3}P_3)=0\, .
\lb{24-0}
\ee
In this equation  the terms containing  $P_3$   do not vanish on-shell. Hence the sum of these terms has  to vanish separately from the other terms. This gives us a relation
for the beta function for the coupling $\lambda_3$ in front of the quartic power of  the Riemann tensor,
\be
\hat{\beta}_3=3v_{1,3}-2\lambda_2\partial_{\lambda_2}v_{1,3}\, .
\lb{25-2}
\ee
We remind that $v_{1,3}$ is linear function of   $\lambda_2$.
The rest of equation (\ref{24-0}) can be resolved  for  a three-loop  beta function for  the metric,
\be
(\beta_3)_{ij}=3 X^{(3)}_{ij}-2\lambda_2\partial_{\lambda_2}X^{(3)}_{ij}-\lambda_2 \left( a\, (P'_2)_{ij}+b\, g_{ij}g^{kl}(P'_2)_{kl} \right)\, ,
\lb{26}
\ee
where  $a$ and $b$ are those that appeared in the one-loop equation (\ref{18}) and $(P'_2)_{ij}$ is a metric variation of invariant $P_2$, see for instance \cite{RRR},
\be
(P'_2)_{ij}=-3R_i^{\ kln}R_{ln}^{\ \ mp}R_{mpjk}+\frac{1}{2}g_{ij} R_{klmn}R^{mn}_{\ \ \ ab}R^{abkl}-6\nabla^k\nabla^l(R_{i kmn}R_{jl}^{\ \ mn})\, ,
\lb{27}
\ee
\be
g^{kl}(P'_2)_{kl}=-  R_{klmn}R^{mn}_{\ \ \ ab}R^{abkl}-\frac{3}{2}\Box (R_{abkl}R^{abkl})-12\nabla^k\nabla^l(R_{kilj}G^{ij}) +O(G^2)\, ,
\lb{28}
\ee
where in (\ref{28})  we used (\ref{25}) and $O(G^2)$ stands for terms quadratic in  $G_{ij}$.

\bigskip

\bigskip
\noindent \underline{\bf $l=4$}

\medskip

In the quartic order (four loops) one finds
\be
&&-4L_{1,4}+L'_0\cdot \beta_4+P_4 \, (\hat{\beta}_4-4v_{1,4})+\lambda_2 P'_2\cdot \beta_2+\lambda_3 P'_3\cdot \beta_1\nonumber \\
&&+(2\lambda_2\partial_{\lambda_2}+3\lambda_3\partial_{\lambda_3})(L_{1,4}+v_{1,4}P_4)=0\, .
\lb{27-1}
\ee
As before, this equation splits on two parts: the first part contain terms that have at least one power  of $G_{ij}$ and the other part contain terms that are due to the Riemann tensor only.
The first part can be used to determine the metric beta function $\beta_4$ while the second part is used to determine the beta function for the higher curvature coupling $\lambda_4$.
In this equation the 3rd term is due to the Riemann tensor only and, thus, it does not vanish on-shell whilst the 1st, 2nd and 5th 
terms contain at least one power of the Einstein tensor $G_{ij}$. The 4th term has  a part due to the Riemann tensor
only and the other part that has at least one power of $G_{ij}$,
\be
&&P'_2\cdot \beta_2=\int 2c_0\left(-\Tr(R^2)\Tr(R^3)-\frac{3}{2}\Tr(R^2)\Box\Tr (R^2)\right)\nonumber \\
&&-\int 24c_0G^{ij}R_{ikjl}\nabla^k\nabla^l\Tr(R^2)+\int 2c_1G^{ij}R_{ikjl}(P'_2)^{kl}+O(G^2)\, ,
\lb{28-1}
\ee
where $(P'_2)_{ij}$ is given by eq.(\ref{27}) and $O(G^2)$ stands for terms quadratic in the Einstein tensor and $\Tr(R^n)$ is the trace of a product of $n$ copies of the  Riemann tensor, $\Tr(R^n)=R^{i_1 i_2}_{\ \ \ i_3 i_4}R^{i_3i_4}_{\ \ \ i_5i_6}\dots R^{i_{2n-1}i_{2n}}_{\ \  \ \ \  \ \ i_1i_2}$. The second line in (\ref{28-1}) contains at least one power of the Einstein tensor and, thus, should be taken into account in (\ref{27-1}) when one determines the four loop beta function for the metric,
\be
&&(\beta_4)_{ij}=4X^{(4)}_{ij}-(2\lambda_2\partial_{\lambda_2}+3\lambda_3\partial_{\lambda_3})X^{(4)}_{ij}+24c_0\lambda_2R_{ikjl}\nabla^k\nabla^l(\Tr R^2) \nonumber \\
&&+6c_1\lambda_2R_{ikjl}R^{kabc}R_{bc}^{\ \ mn}R^l_{\ amn}-12c_1\lambda_2R_{ikjl}\nabla^a\nabla^b(R^k_{\ amn}R_{b}^{\  l mn})\nonumber \\
&&-\lambda_3(a(P'_3)_{ij}+bg^{kl}(P'_3)_{kl})+O(G)\, ,
\lb{29}
\ee
where $O(G)$ are terms linear in $G_{ij}$.

There is more than one invariant of 5th order that can be constructed from  the Riemann tensor. Here we list some of such invariants,
\be
P_4^{(1)}=\int \Tr(R^2)\Tr(R^3)\, , \ \ \  P_4^{(2)}=\int \Tr R^2~\Box\Tr R^2\, , \ \ \  P_4^{(3)}=\int \Tr (R^5)\, , \ \ \  \dots \, .
\lb{30}
\ee
Respectively, the coupling constant has an extra index $\lambda_4^{(a)}\, , \ a=1,2,\dots$. From eq.(\ref{27-1}) we find the respective beta functions
\be
&& \hat{ \beta}_4^{(1)}=4v_{1,4}^{(1)}-(2\lambda_2\partial_{\lambda_2}+3\lambda_3\partial_{\lambda_3})v_{1,4}^{(1)}+2\lambda_2 c_0\, ,   \nonumber \\
&& \hat{\beta}_4^{(2)}=4v_{1,4}^{(2)}-(2\lambda_2\partial_{\lambda_2}+3\lambda_3\partial_{\lambda_3})v_{1,4}^{(2)}+3\lambda_2 c_0\, , \nonumber \\
&&  \hat{\beta}_4^{(a)}=4v_{1,4}^{(a)}-(2\lambda_2\partial_{\lambda_2}+3\lambda_3\partial_{\lambda_3})v_{1,4}^{(a)}\, , \ a=3,\, \dots 
 \lb{31}
 \ee

 \section{The higher pole counterterms}
 \setcounter{equation}0
\bigskip

Now it is time to look at the higher poles in the RG equation (\ref{13}). The vanishing condition for  the coefficient in front of the pole $1/\epsilon^k\, , \ k\geq 1$ gives us equation,
\be
&&-L_{k+1}-V_{k+1}+L'_{k+1}\cdot g+V'_{k+1}\cdot g+\sum_{p=2}p\lambda_p\partial_{\lambda_p}(L_{k+1}+V_{k+1})\nonumber \\
&&+L'_k\cdot \beta+V'_k\cdot \beta+\sum_{p=2}\hat{\beta}_p\partial_{\lambda_p}(L_k+V_k)=0\, , \ \ k\geq 1
\lb{32}
\ee
This is a recurrence relation that can be used to determine the counter-terms $L_{k+1}$ and $V_{k+1}$ provided the lower  pole counter-terms $L_{p}\, , \ V_{p}\, , \ p=k\, ,   k-1\, , \dots$ are given.
In order to  start  the recurrence procedure   one has to know  the single pole terms $L_1$ and $V_{1}$.

Expanding in powers of the Newton constant as in (\ref{9}) and (\ref{9-1}) we find the recurrence relation for terms $L_{k,l}$ and $V_{k,l}$ that appear in this expansion,
\be
&&(1-\frac{1}{l}\sum_{p=2}^{l-1}p\lambda_p\partial_{\lambda_p})(L_{k+1,l}+V_{k+1,l})\nonumber \\
&&=\frac{1}{l}\sum_{p=1}^{l-k}(L'_{k,l-p}+V'_{k,l-p})\cdot \beta_p+\frac{1}{l}\sum_{p=2}^{l-1}\hat{\beta}_p\partial_{\lambda_p}(L_{k,l}+V_{k,l})\, , \ \ \ l\geq k+1
\lb{33}
\ee
where $\beta_p$ are the coefficients in the power series (with respect to $G_N$) of the metric beta function and $\hat{\beta}_p$ is the beta function for a higher curvature coupling $\lambda_p$.

In the effective action  the expansion goes in two directions:  the powers series in $1/\epsilon$ and the powers of the Newton constant $G_N$.
Interchanging the order of these two expansions one finds
\be
\sum_{k=1}\frac{1}{\epsilon^k}\sum_{l=k}G_N^{l-1}(L_{k,l}+V_{k,l})=\sum_{l=1}G_N^{l-1}\sum_{k=1}^l \frac{1}{\epsilon^{k}}(L_{k,l}+V_{k,l})
\lb{34}
\ee
The expansion on the right hand side is the loop expansion that indicates that in $l$-th loop the UV divergences run from a single pole $k=1$ to the highest pole $k=l$.
We stress once  again that the basic information is always contained in the single pole.   The higher poles are expressed in terms of the single pole using equation (\ref{33}).

\medskip

Below in this section we analyze the solutions of Eq.(\ref{33}) for certain values of $l$ and $k$.

\subsection{$l=2$}

In this case there is no dependance on the couplings $\lambda_p$ and the RG equation (\ref{33}) takes a rather simple form
\be
L_{2,2}+V_{2,2}=\frac{1}{2}L'_{1,1}\cdot \beta_1\, ,
\lb{36}
\ee
where we take into account that $V_{1,1}=0$. In this equation the right hand side vanishes on-shell since $\beta_1$ is linear in the Einstein tensor. On the left hand side of this
equation $L_{2,2}$ also vanishes since by assumption $L_{k,l}$ contain at least one power of $G_{ij}$. Since $V_{2,2}$ is the only term in (\ref{36})
that does not contain the Ricci tensor or the Ricci scalar  it  has to vanish identically,
\be
V_{2,2}=0\, \ (v_{2,2}=0)\, .
\lb{37}
\ee
We remark that the two loop result  (\ref{37})   was first obtained, using methods different from ours, by Chase in 1982 \cite{Chase:1982sf} (see also discussion in \cite{Marcus:1984ei}).
The counter-term $L_{1,1}$ takes the form (\ref{18}). Using the beta function (\ref{20}) we find
\be
&&L_{2,2}=\int (a^2R_{injl}G^{nl}G^{ij}+\frac{a^2}{2}G_{ij}\Box G^{ij}-\frac{1}{2}(a^2+6b^2+4ab)G\Box G\nonumber \\
&&-(\frac{3}{4}a^2+ab)GG^2_{ij}+\frac{1}{4}(ab+ a^2)G^3)\, .
\lb{40}
\ee
We see that $L_{2,2}$ is  at least quadratic in $G_{ij}$. 

\subsection{$l=3$}
Two values of $k$ are possible: $k=1$ and $k=2$. For $k=1$ the RG equation (\ref{33}) is
\be
(1-\frac{2}{3}\lambda_2\partial_{\lambda_2})(L_{2,3}+V_{2,3})=\frac{1}{3}(L'_{1,2}+V'_{1,2})\cdot \beta_1+\frac{1}{3}L'_{1,1}\cdot \beta_2+\frac{1}{3}\hat{\beta}_2\partial_{\lambda_2}(L_{1,3}+V_{1,3}))
\lb{xx}
\ee
while for $k=2$ the RG equation is
\be
(1-\frac{2}{3}\lambda_2\partial_{\lambda_2})(L_{3,3}+V_{3,3})=\frac{1}{3}L'_{2,2}\cdot \beta_1+\frac{1}{3}\hat{\beta}_2\partial_{\lambda_2}(L_{2,3}+V_{2,3}))\, ,
\lb{xxx}
\ee
where, as we have shown earlier, $\hat{\beta}_2=2v_{1,2}$. In this order the counter-terms can be at most linear in $\lambda_2$ so that $V_{1,3}=V^{(0)}_{1,3}+V^{(2)}_{1,3}\lambda_2$ and the same for
the counter-terms $L_{1,3}$. 
Here and below we  use the notations for a linear function of $\lambda$: $f(\lambda)=f^{(0)}+\sum_{a=2}f^{(a)}\lambda_a$. 
Separating in each equation the terms vanishing on-shell and the terms non-vanishing on-shell 
one solves these two equations and obtains
\be
&&V_{2,3}=\frac{2}{3}v_{1,2}V^{(2)}_{1,3}\, , \nonumber \\
&&L_{2,3}=\frac{2}{3}v_{1,2}L^{(2)}_{1,3}+\frac{1}{3}(L'_{1,2}+V'_{1,2})\cdot\beta_1+\frac{1}{3}L'_{1,1}\cdot \beta_2\, ,\nonumber \\
&&V_{3,3}=0\ (v_{3,3}=0)\, , \ \ L_{3,3}=\frac{1}{3}L'_{2,2}\cdot \beta_1\, .
\lb{y}
\ee
Few things we should notice. First of all, none of the counter-terms $V_{k,3}\, , \ k=2\, , \, 3$ and $L_{k,3}\, , \ k=2\, , \, 3$ depends on $\lambda_2$.  Next, $V_{3,3}$ vanishes identically 
in exact parallel with the vanishing of $V_{2,2}$ and $V_{1,1}$. Finally, looking a bit more careful at $L_{3,3}$ we can see that it is  at least quadratic in the Einstein tensor, similarly to $L_{2,2}$ (\ref{40}) and $L_{1,1}$ (\ref{18}).
We will see whether some of these observations persist to a higher loop order.

\subsection{$l=4$}
In this case three values of $k$ are possible: $k=1$, $k=2$ and $k=3$. For the beta function $\hat{\beta}$ one has that $\hat{\beta}_2=2v_{1,2}$ is independent of $\lambda$
and that $\hat{\beta}_3=3v^{(0)}_{1,3}+v^{(2)}_{1,3}\lambda_2$ is linear function of $\lambda_2$.  Here $v_{1,3}=v^{(0)}_{1,3}+v^{(2)}_{1,3}\lambda_2$.
The metric beta functions $\beta_1$ and $\beta_2$ do not depend on $\lambda$ while $\beta_3=\beta_3^{(0)}+\beta_3^{(2)}\lambda_2$ is linear in $\lambda_2$ (see (\ref{26})). The analysis of the RG equations (\ref{33}) goes along same line as for $l=3$. We skip the details of the analysis and below summarize the results.

\bigskip

\noindent\underline{$k=1$}
\be
&&V^{(0)}_{2,4}=\widehat{V}_{2,4}  + \frac{v_{1,2}}{2}V^{(2)}_{1,4}+\frac{3}{4}v^{(0)}_{1,3}V^{(3)}_{1,4}\, , \  \  \widehat{V}_{2,4}=\frac{1}{4}(L'_{1,2}+V'_{1,2})\cdot \beta_2|_{G_{ij}=0}\, ,\nonumber \\
&&L^{(0)}_{2,4}=\frac{1}{4}(L^{(0)}_{1,3}+V^{(0)}_{1,3})'\cdot \beta_1+\widehat{L}_{2,4}+\frac{1}{4}(L^{}_{1,1})'\cdot \beta^{(0)}_3+\frac{1}{2}v_{1,2}L^{(2)}_{1,4}+\frac{3}{4}v^{(0)}_{1,3}L^{(3)}_{1,4}\, ,\nonumber \\
&&\widehat{L}_{2,4}=\frac{1}{4}(L'_{1,2}+V'_{1,2})\cdot \beta_2-\widehat{V}_{2,4}\, , \ \ \  V^{(2)}_{2,4}=\frac{1}{2}v^{(2)}_{1,3}V^{(3)}_{1,4}\, , \nonumber \\
&&L^{(2)}_{2,4}=\frac{1}{2}(L^{(2)}_{1,3}+V^{(2)}_{1,3})'\cdot \beta_1+\frac{1}{2}L'_{1,1}\cdot \beta^{(2)}_{3}+\frac{1}{2}v^{(2)}_{1,3}L^{(3)}_{1,4}\, ,\nonumber \\
&&V^{(3)}_{2,4}=L^{(3)}_{2,4}=0\, .
\lb{4-1}
\ee

\bigskip

\noindent\underline{$k=2$}
\be
&&V^{(0)}_{3,4}=\frac{1}{2}v_{1,2}V^{(2)}_{2,4}\, , \nonumber \\
&&L^{(0)}_{3,4}=\frac{1}{4}(L'_{2,3}+V'_{2,3})\cdot \beta_1+\frac{1}{4}L'_{2,2}\cdot \beta_2+\frac{1}{2}v_{1,2}L^{(2)}_{2,4}\, ,\nonumber \\
&&V^{(2)}_{3,4}=L^{(2)}_{3,4}=0\, ,  \ \ \  V^{(3)}_{3,4}=L^{(3)}_{3,4}=0\, .
\lb{4-2}
\ee

\bigskip

\noindent\underline{$k=3$}
\be
V_{4,4}=0\, , \ \ L_{4,4}=\frac{1}{4}L'_{3,3}\cdot \beta_1\, .
\lb{4-3}
\ee
The counter-terms (\ref{4-2}) and (\ref{4-3}) do not depend on $\lambda_2$ or $\lambda_3$.  Counter-term $V_{4,4}$ vanishes identically similarly to $V_{3,3}$ and $V_{2,2}$.
A careful analysis (which we perform in the next section) demonstrates that $L_{4,4}$ is a polynomial in  $G_{ij}$ that starts with a quadratic term. This is similar to
what we have found for $L_{2,2}$ and $L_{3,3}$ and what was known  for $L_{1,1}$ (\ref{18}).

\section{The GR counter-terms}
\setcounter{equation}0
\bigskip

As we have seen above, the counter-terms have different origins. Some of them originate from the Feynman diagrams with only the usual GR vertices, that come from the General Relativity action, and the other counter-terms come from the diagrams where 
additional vertices due to the higher curvature couplings are present. In this section our goal is to isolate those counter-terms that are due to the GR vertices only. We call them the GR counter-terms.
They can be obtained from the total counter-terms $V_{k,l}$ and $L_{k,l}$ by taking the limit of vanishing couplings $\{\lambda_p\}$ and neglecting derivatives of the total counter-terms with respect to $\lambda_p\, , p\geq 2$.
We  will denote  the GR counter-terms vanishing on-shell  as ${\cal L}_{k,l}$  and the counter-terms non-vanishing on-shell  as ${\cal V}_{k,l}$. The recurrence relations for the GR counter-terms are obtained from (\ref{33})
in the limiting procedure just described,
\be
{\cal{ L}}_{k+1,l}+{\cal{ V}}_{k+1,l}=\frac{1}{l}\sum_{p=1}^{l-k}({\cal{ L}}'_{k,l-p}+{\cal{ V}}'_{k,l-p})\cdot  \beta^{(0)}_p\, , \ \ \ l\geq k+1\, 
\lb{G1}
\ee
where $\beta^{(0)}_p$ is the metric beta function in the limit of vanishing $\lambda$. For $p=1$ and $p=2$ it is the same as $\beta_1$ and $\beta_2$.
As one can see from our analysis in the previous section $V_{k,k}={\cal V}_{k,k}$ and $L_{k,k}={\cal L}_{k,k}$ for $k=1\, , \ 2\, , \  3\, , \ 4$.

\subsection{Some general properties of GR counter-terms in  the highest pole $k=l$}

The GR counter-terms in the  highest pole   satisfy equation
\be
{\cal L}_{k+1,k+1}+{\cal V}_{k+1,k+1}=\frac{1}{k+1}({\cal L}'_{k,k}\cdot \beta_1+{\cal V}'_{k,k}\cdot \beta_1)\, , \ k\geq 1
\lb{35}
\ee
where $\beta_1$ is the metric beta function in one loop (\ref{20}). The right hand side of this equation necessarily contains at least one power of $G_{ij}$ (due to $\beta_1$) so does ${\cal L}_{k+1,k+1}$ in the left hand side  of this equation.
The only term that does not contain $G_{ij}$ is ${\cal V}_{k+1,k+1}$ and, hence, it has to vanish,
 \be
 {\cal V}_{k+1,k+1}= 0\, , \ \ k\geq 1\, .
\lb{38}
\ee
So that one has  the following

\medskip

\noindent{\it \underline{Statement 1.}  The GR counter-terms in the highest  pole $k=l$ at any given loop order $l$  vanish on-shell.}

\bigskip

\noindent  Taking into account (\ref{38}) the  RG equation (\ref{35}) can be written in a simpler form

\be
{\cal L}_{k+1,k+1}= \frac{1}{k+1}{\cal L}'_{k,k}\cdot \beta_1\, .
\lb{39}
\ee
We have seen that $L_{2,2}$  and $L_{3,3}$ vanish   quadratically in $G_{ij}$.  This property can be extended for the GR counter-terms  ${\cal L}_{k,k}$ for any $k\geq 2$.
This can be done using the induction.  The counter-term $L_{2,2}$ is at least quadratic in $G_{ij}$. Let us assume that ${\cal L}_{k,k}\, \ k>2$ is at least quadratic in $G_{ij}$. So that it can be
represented in the form
\be
{\cal L}_{k,k}=\int G^{ij}Y_{ij}^{(k)}\, ,
\lb{40-1}
\ee
where $Y^{ij}_{(k)}$ has a term  linear in the Einstein tensor $G$. Then, varying (\ref{40-1}) with respect to metric and neglecting the terms quadratic  and of a higher order in $G_{ij}$ one has
$$
({\cal L}'_{k,k})_{ij}=-2R_{i\ j}^{\ a \ b}Y^{(k)}_{ab}+ \nabla_i\nabla^a Y^{(k)}_{aj}+\nabla_j\nabla^a Y^{(k)}_{ai}-g_{ij}\nabla^a\nabla^b Y^{(k)}_{ab}-\Box Y^{(k)}_{ab}-\nabla_i\nabla_j Y^{(k)}+g_{ij}\Box Y^{(k)}\, ,
$$
where $Y^{(k)}=g^{ij}Y^{(k)}_{ij}$.
Substituting this into eq.(\ref{39}) one finds that ${\cal L}_{k+1,k+1}$ can be represented in a form similar to (\ref{40-1}) with
$$
Y^{(k+1)}_{ij}= \frac{1}{k+1}(-2aR_{iajb}Y^{ab}_{(k)}+ag_{ij}\Box Y^{(k)}-a\Box  Y^{(k)}_{ij}-(a+2b)g_{ij}(\nabla^a\nabla^b Y^{(k)}_{ab}-\Box Y^{(k)}))+O(G^2)
$$
that demonstrates that $Y^{(k+1)}_{ij}$  as well has necessarily a term   linear in the Einstein tensor. Hence, one concludes that  ${\cal L}_{k+1,k+1}$ is at least  quadratic in $G_{ij}$, i.e.
\be
{\cal L}_{k+1,k+1}= O(G^2)\, , \ \  k\geq 1
\lb{40-4}
\ee
Thus, one has   the following

\medskip

\noindent{\it \underline{Statement 2.}  At any loop order $l$ the GR counter-terms in the highest pole $k=l$   vanish  on-shell quadratically.}

\bigskip

\noindent We remark here that both $a$ and $b$ that appear in the one-loop counter-term $L_{1,1}$ (\ref{18}) depend on the gauge conditions, see \cite{Kallosh:1978wt} and \cite{Barvinsky:1985an}.
There may exist a certain gauge for which  both $a=0$ and $b=0$, see  \cite{Kallosh:1978wt} where some of such gauge conditions were found. In this case all the  GR counter-terms in the highest poles vanish identically, ${\cal L}_{k,k}= 0\, , \ k\geq 1$
(as well as the one-loop beta function $\beta_{1,ij}$). This is so up to a topological Euler term which we 
ignore here.  On the other hand, in an alternative approach which makes use of the so-called unique quantum effective action of Vilkovisky one ends up  with the certain non-vanishing values of $a$ and $b$
which are claimed to be unique in quantum gravity \cite{Barvinsky:1984jd}, \cite{Fradkin:1983nw}. In any case, it makes sense to  keep the consideration general  and consider the  arbitrary non-vanishing $a$ and $b$. Then our Statement 2 is non-trivial since it restricts the possible dependence of the
highest pole counter-terms  on the Einstein tensor. For instance, a possibility that ${\cal L}_{k,k}$ may vanish by  a power law  $O(G^k)$ depending on the value of $k$   is ruled out by Statement 2.
Among other things, our results may also serve as a consistency condition to be used as a tool to  check the higher loop calculations.

\subsection{Some general properties of GR counter-terms in a sub-leading pole $k=l-1$}

Consider now the first sub-leading  pole at a given loop order. From the recurrence relation  (\ref{33})
we find
\be
{\cal L}_{k+1,k+2}+{\cal V}_{k+1,k+2}= \frac{1}{k+2}\left(({\cal L}'_{k,k+1}+{\cal V}'_{k,k+1})\cdot\beta_1+{\cal L}'_{k,k}\cdot\beta_2\right)\, ,
\lb{41}
\ee
where we already took into account (\ref{38}). Beta function $\beta_2$ is given by Eqs.(\ref{23})-(\ref{24}). It does not necessarily vanish on-shell.  However, by our Statement 2, ${\cal L}'_{k,k}$ vanishes linearly in $G_{ij}$.
Therefore, all terms on the r.h.s of (\ref{41}) vanish on-shell.  On the l.h.s. of (\ref{41}) ${\cal L}_{k+1,k+2}$ vanishes at least linearly in $G_{ij}$.
Thus, we see that in eq. (\ref{41}) there is only one term, ${\cal V}_{k+1,k+2}$, that does not vanish on-shell. And hence it has to be zero,
\be
{\cal V}_{k+1,k+2}= 0\, , \ \ \ k\geq 1
\lb{42}
\ee

\medskip

\noindent{\it \underline{Statement 3.}  At any loop order $l$ the  GR counter-terms in the first sub-leading pole $k=l-1$ do not contain any terms that are due to the Riemann tensor only.}

\bigskip

We note that (\ref{42}) is valid starting with $k=1$ and, thus, is not  true for $k=0$. Indeed, $V_{1,2}=v_{1,2}P_2$ is a single pole that appears in two loops. It is proportional to cubic invariant $P_2$ (\ref{22-1}).
Let us consider (\ref{41}) for $k=1$,
\be
{\cal L}_{2,3}= \frac{1}{3}(L'_{1,2}+V'_{1,2})\cdot\beta_1+L'_{1,1}\cdot \beta_2\, .
\lb{43}
\ee
It is an expression for a second order pole that appears in three loops.
 We can not claim that $L_{2,3}$  is quadratic in $G_{ij}$. Indeed,   on the r.h.s. $V'_{1,2}=v_{1,2}P'_2$ is given by  (\ref{27}) that is
cubic in the Riemann tensor. Taking that $\beta_{1}$ is linear in $G_{ij}$ we conclude that $L_{2,3}$ necessarily contains a term of the form $G_{ij}({\cal R}^3)^{ij}$, where $\cal R$ stands for the Riemann tensor.
Similar reasonings are valid for the 3rd term in the r.h.s. of (\ref{43}). The direct calculation gives us the following expression,
\be
&&{\cal L}_{2,3}=\int (-av_{1,2} G_{ij}R_{i}^{\ kln}R_{ln}^{\ \ mp}R_{mpjk}+v_{1,2}(\frac{a}{6}-\frac{b}{3})G\Tr R^3\nonumber \\
&& -2av_{1,2}G_{ij}\nabla^k\nabla^n(R_{ikmn}R_{jl}^{\ \ mn})-(\frac{b}{2}v_{1,2}+c_0(3b+a))G\Box\Tr R^2+O(G^2))\, .
\lb{44}
\ee
For a  larger value of $k\geq 3$ equation (\ref{41}), provided one uses (\ref{42}),  reduces to
\be
{\cal L}_{k+1,k+2}=\frac{1}{k+2}\left({\cal L}'_{k,k+1}\cdot\beta_1+{\cal L}'_{k,k}\cdot\beta_2\right)\, .
\lb{45}
\ee
Clearly, in the r.h.s. of this equation one always has a term that vanishes  linearly in $G_{ij}$. 
This, in particular, rules out  the possibility for ${\cal L}_{k,k+1}$ to vanish by a power law with the power growing with  $k$.

\subsection{ Some general properties of GR counter-terms in a sub-leading pole $k=l-2$}

The recurrence relation (\ref{33}) in this order leads to equation
\be
{\cal L}_{k+1, k+3}+{\cal V}_{k+1,k+3}=\frac{1}{k+3}(({\cal L}'_{k,k+2}+{\cal V}'_{k,k+2})\cdot \beta_1+({\cal L}'_{k,k+1}+{\cal V}'_{k,k+1})\cdot\beta_2+{\cal L}'_{k,k}\cdot \beta^{(0)}_3)\, ,
\lb{46}
\ee
where we already took into account that ${\cal V}_{k,k}=0\, , \ k\geq 1$. We have  shown  earlier that ${\cal V}_{k,k+1}=0\, , \ k\geq 2$. However, for $k=1$ it is non-zero since ${\cal V}_{1,2}$ is a non-trivial single pole.
Looking at equation (\ref{46}) we note that $\beta_1$ is linear in $G_{ij}$, ${\cal L}_{k,k}$ is quadratic in $G_{ij}$ and, hence, ${\cal L}'_{k,k}$ is at least linear in $G_{ij}$. 
On the other hand, ${\cal L}_{k,k+1}$ is linear in $G_{ij}$ and hence ${\cal L}'_{k,k+1}$ may be non-vanishing if $G_{ij}=0$.
Putting $G_{ij}=0$ on both sides of (\ref{46})
we find that
\be
{\cal V}_{k+1,k+3}=\frac{1}{k+3}{\cal L}'_{k,k+1}\cdot \beta_2|_{G_{ij}=0}\, , \  k\geq 2\, .
\lb{47}
\ee
This indicates that  the second sub-leading pole  may not vanish on-shell.  Considering the decreasing order of the pole at a fixed loop order $l$, this is the first time when a higher pole may not vanish on-shell.
The value $k=1$ is a special case. One has more non-vanishing terms in equation (\ref{46}) in this case,
\be
{\cal L}_{2,4}+{\cal V}_{2,4}=\frac{1}{4}(({\cal L}'_{1,3}+{\cal V}'_{1,3})\cdot \beta_1+({ L}'_{1,2}+{V}'_{1,2})\cdot \beta_2+{L}'_{1,1}\cdot \beta^{(0)}_3)\, .
\lb{48}
\ee
The terms in the r.h.s. of this equation are due to a single pole $(k=1)$. We recall that $L_{1,1}$ is quadratic in $G_{ij}$ and hence $L'_{1,1}$ is linear in $G_{ij}$.
Putting $G_{ij}$ on both sides of (\ref{48}) we find 
\be
{\cal V}_{2,4}=\frac{1}{4}(L'_{1,2}+V'_{1,2})\cdot \beta_2|_{G_{ij}=0}\, \ .
\lb{49}
\ee
This expression is what we called  $\widehat{V}_{2,4}$ in (\ref{4-1}).
A general form for the beta function $\beta_{2,ij}$ is given by (\ref{23}), (\ref{24}). Imposing $G_{ij}=0$ only the term with $c_0$ survives in (\ref{24}). This term is proportional to $g_{ij}$ so that (\ref{49})
is simplified in the limit $G_{ij}=0$,
\be
{\cal V}_{2,4}=\frac{1}{2}c_0(L'_{1,2}+V'_{1,2})\cdot g  \Tr R^2|_{G_{ij}=0}\, .
\lb{50}
\ee
$L_{1,2}$ has the form (\ref{22}), (\ref{24}) while $V_{1,2}=v_{1,2}P_2$, whose metric variation is given by (\ref{27}), (\ref{28}). Putting everything together we find
\be
{\cal V}_{2,4}=\frac{3}{2}c_0(c_0-\frac{1}{2}v_{1,2})\int \Tr R^2 \Box \Tr R^2-\frac{1}{2}c_0v_{1,2}\int \Tr R^2\Tr R^3\, .
\lb{51}
\ee

A similar analysis can be done for $k\geq 2$, see equation (\ref{47}). Since ${\cal L}_{k,k+1}$ is linear in $G_{ij}$ it can be written as follows
\be
{\cal L}_{k,k+1}=\int G_{ij}Z_{(k)}^{ij}+O(G^2)\, ,
\lb{52}
\ee
where tensor  $Z^{ij}_{(k)}$ has a  curvature order $k$, it is  constructed from the Riemann tensor so that it does not vanish on-shell.  Then one finds that
\be
({\cal L}'_{k,k+1})_{ij}g^{ij}=-\nabla_\alpha\nabla_\beta Z^{\alpha\beta}_{(k)}+\Box Z^{\alpha\beta}_{(k)}g_{\alpha\beta}+O(G)\, .
\lb{53}
\ee
Using this equation and (\ref{23}), (\ref{24}) for the beta function $\beta_2$ one finds 
\be
{\cal V}_{k+1,k+3}=\frac{2c_0}{k+3}\int (-\nabla_\alpha\nabla_\beta Z^{\alpha\beta}_{(k)}+\Box Z^{\alpha\beta}_{(k)}g_{\alpha\beta})\Tr R^2\, .
\lb{54}
\ee
Our analysis in this subsection can be summarized in the following

\medskip

\noindent{\it \underline{Statement 4.} At any loop order $l\geq 4$ the GR counter-terms in a sub-leading pole $k=l-2$ do not necessarily vanish on-shell: there may appear
some terms that are due to the Riemann tensor only.}

\medskip

The appearance of the non-vanishing on-shell terms is possible in other sub-leading poles, $k=l-3\, , \ l-4\, , \dots\, .$  This can be easily analyzed using our equation (\ref{G1}). 
We, however, do not consider this question in the present paper.

\subsection{Remarks on previous works}

We finish this section with some remarks concerning the compatibility of our work with the previous results in the literature. 
The only earlier paper, we are aware of,  that actually computed  the higher pole counter-terms in quantum gravity is a paper of Goroff and Sagnotti \cite{Goroff:1985th}.
In equation (3.18) of this paper they presented the result of an off-shell  2-loop calculation.  It contains both the single pole terms and the double poles, i.e. in our notations they computed $V_{1,2}$, $L_{1,2}$ and 
$V_{2,2}$, $L_{2,2}$. We are here interested in the double pole counter-terms.  Unfortunately, the comparison of these earlier results with our analysis  shows  certain signs of disagreement.
Indeed, in (3.18) of \cite{Goroff:1985th}  the double pole includes terms cubic in the Ricci tensor, that are absent in our eq.(\ref{40}). Even more, in (3.18) there  presents a term which is linear in $G_{ij}$,
$R_{\alpha\beta\gamma\delta}R^{\alpha\beta\gamma\sigma}R^\delta_{\ \sigma}$, 
although in our analysis the counter-term $L_{2,2}$ is necessarily quadratic in $G_{ij}$. We believe that a possible source of the disagreement is  the following.  The authors of  \cite{Goroff:1985th}, using a weak field approximation over Minkowski
spacetime, compute in two loops the UV divergences for the cubic vertices that contain six derivatives and the result  of the calculation then is used to fix the coefficients in front of the possible cubic curvature invariants.
They count nine cubic invariants (see their equations (3.12 a-c)): $I_1=R\Box R$, $I_2=R^3$, $I_3=R_{ij}\Box R_{ij}$, $I_4=RR_{ij}^2$, $I_5=R_{ik}R_{jl}R^{ijkl}$, $I_6=R_i^{\ j}R_{j}^{\ k}R_{k}^{\ i}$,
$I_7=RR_{ijkl}R^{ijkl}$, $I_{8}=R_{ij}R^{iklm}R^j_{\ klm}$, $I_9=R_{ij}^{\ \ kl}R_{kl}^{\ \ mp}R_{mp}^{\ \ ij}$. In their calculation the authors of \cite{Goroff:1985th} drop the trace of the metric perturbation and its divergences.
In this way they apparently can not determine the coefficients in front of invariants that contain the Ricci scalar. They, however, say that they can determine the coefficients for five invariants $I_3$, $I_5$, $I_6$, $I_8$ and $I_9$ that do not contain the Ricci scalar.

It appears  that the authors of  \cite{Goroff:1985th} were not aware of the relation (\ref{25}). Using this relation one finds that invariant $I_8$ is not an independent invariant,
\be
I_8=\frac{1}{4}(I_2+I_7)+2(-I_4+I_5+I_6)\, .
\lb{55}
\ee
So that, in reality, there are only 8 independent, cubic in curvature, invariants   and not 9 as was assumed in \cite{Goroff:1985th}.  Invariant $I_8$ has to be excluded.
This means that, if everything were consistent, one would have been  able to fix the coefficients for four (not five!) invariants that did not contain the Ricci scalar.
This of course would correct the  numerical factors   in (3.18) of \cite{Goroff:1985th}.

The other related earlier work that discusses (but does not compute) the higher pole counter-terms in quantum gravity is \cite{Kazakov:1987ej}. It was assumed in \cite{Kazakov:1987ej} that the GR highest pole, what we call 
${\cal L}_{l,l}$, has a first order zero, i.e. vanishes on-shell linearly. This was important for that the renormalization scheme suggested in \cite{Kazakov:1987ej} actually worked. As  we demonstrate this here
 the counter-terms ${\cal L}_{l,l}$ vanish quadratically so that the scenario suggested in \cite{Kazakov:1987ej} can not be realized in pure quantum gravity (with zero cosmological constant).

\section{Quantum action as a renormalized  gravitational action}
\setcounter{equation}0
\bigskip

In this paper we have introduced two renormalization group equations: one for the metric  (\ref{2}) and the other for the quantum effective action (\ref{12}). These two sets of equations
appear to know about each other through the metric beta function $\beta_{ij}$: it is determined by the single pole terms in the effective action by means of the equation (\ref{15}), (\ref{16}).
In this section we want to show that there is a deeper relation between the two RG equations. This relation can  in fact  be anticipated taking into account 
the way the renormalization works  in the case of  the renormalizable field theories.
Indeed, in a renormalizable field theory all UV divergences in the quantum effective action can be hidden in the field renormalization and the renormalization of the coupling constants so that
the quantum action takes the original classical (bare) form if expressed in terms of the bare fields and couplings. 
In the case of gravity the classical (bare) action  should include not only the original General Relativity term $G_N^{-1}L_0$ but also the higher curvature terms  $W=\sum_{l=2}G_N^{l-1}\lambda_l P_l$
that are needed to renormalize the UV divergent terms $V_{1}=\sum_{l=2}v_{1,l}P_l$.  The analysis in this section we do in two steps. First, we demonstrate that  the renormalization of the bare GR action
correctly reproduces a certain class of the UV divergent terms that we specify below.    This is true not only for the single pole terms but also for the higher order poles.
For the sake of completeness, the latter agreement we check in detail for the double pole.  It is of course guaranteed by the RG structure that the higher poles agree provided the single pole is the same.
Then, in the second part of this section we make a general statement that all UV divergent terms can be hidden in the renormalization of the total gravitational  action, $L_{\rm gr}=G^{-1}_NL_0+W$.
There we focus only on the single pole terms.

\subsection{Renormalization of the GR action}

As we have explained above we first concentrate our attention on the terms that are independent of the higher curvature couplings.
Therefore, in  our analysis in this section we assume that such terms are not present, i.e. consider the case when $v_{1,l}=0$ and $\lambda_l=0$. Since the quantum action depends on $v_{1,l}$ and $\lambda_l$ analytically this limit can always be arranged. Then for the corresponding part in the  quantum action (\ref{12}) one finds
\be
\overline{L}_Q(g_R)=L_Q(g_R, \lambda_l=0, v_{1,l}=0)=\mu^{-\epsilon}(\frac{1}{G_N}L_0(g_R)+\sum_{k=1}\frac{1}{\epsilon^k}({ L}_k(g_R)+{V}_k(g_R)))\, ,
\lb{60}
\ee
where ${V}_1(g_R)=0$ as we just have explained. Respectively, in the expression for the metric  beta function one has to put $v_{1,l}=0$ and $\lambda_l=0$.

We now want to show that (\ref{60}) is in fact identical to the classical GR gravitational action expressed in terms of the bare metric according to (\ref{2}).
Thus, our claim is that
\be
\overline{L}_Q(g_R)=\frac{1}{G_N} L_0(g_B(g_R))\, , \ \  L_0=-\int d^4x \sqrt{g}R\, .
\lb{62-0}
\ee
So that all divergences that are present in (\ref{60}) are, effectively, hidden in the classical gravitational action.
Provided $g_B(g_R)$ takes the form (\ref{2}) one has that
\be
\frac{1}{G_N} L_0(g_B(g_R))=\frac{\mu^{-\epsilon}}{G_N}L_0(g_R+\sum_{k=1}\epsilon^{-k}h_{(k)}(g_R))\, .
\lb{61}
\ee
The latter expression can be expanded as a formal power series in $\epsilon^{-1}$,
\be
\frac{1}{G_N} L_0(g_B(g_R))=\frac{\mu^{-\epsilon}}{G_N}(L_0+L_0'\cdot h_1\epsilon^{-1}+\frac{1}{2}((L_0''\times h_1)\cdot h_1+2L_0'\cdot h_2)\epsilon^{-2}+\dots)\, ,
\lb{62}
\ee
where we use the definitions  (\ref{4-11})  and (\ref{0}).
The expansion (\ref{62}) is an appropriate generalization of Fa\`a di Brunno's formula for derivatives of  a composite function. A simple form of it is expressed in terms of the Bell polynomials.
Here we  concentrate our attention only on  the first few terms in this formula.
 
 Comparing the expressions (\ref{60}) and (\ref{62}) we see that the first term in both expressions is the same, ${G^{-1}_N} L_0$. So that let us look at the second term in both expressions.
 In equation (\ref{62}) this term is
 \be
 \frac{1}{G_N}\epsilon^{-1} L'_0\cdot h_1=\epsilon^{-1}\sum_{l=1}G^{l-1}_N   L'_0\cdot h_{1,l}=\epsilon^{-1}\sum_{l=1}G^{l-1}_N \, l^{-1}L'_0\cdot \beta_l   \, ,
 \lb{63}
 \ee
 where we used the expansion in powers of $G_N$, $h_1=\sum_{l=1}G^l_N h_{1,l}\, $, and in the last equality we used the relation (\ref{8}) between the metric beta function  $\beta_l$ ($\beta=\sum_{l=1}G^l_N\beta_l$)
 and $h_{1,l}$: $h_{1,l}=l^{-1}\beta_{l}$. The next point is that eq.(\ref{16}), in the limit we consider in this section, reduces to equation 
 \be
 L_{1,l}=l^{-1}L'_0\cdot \beta_l=L'_0\cdot h_{1,l}\, .
 \lb{64}
 \ee
So that eq.(\ref{63}) is precisely $\epsilon^{-1}L_1=\epsilon^{-1}\sum_{l=1}G^{l-1}_N\, L_{1,l}$. We remind once again that $V_{1}=0$ in the limit we consider here so that only $L_1$ appears as a single pole.
We thus have proved that the single pole terms in both expressions (\ref{60}) and (\ref{62}) are identical. Since the higher poles are determined by the single pole and the metric beta function
(that is by itself related to the single pole and is, thus,  the same in both cases) this is sufficient for proving that all other terms (higher poles) in (\ref{60}) and (\ref{62}) are the same and the  two expressions are identical.
Below we check the equality for the double pole $\epsilon^{-2}$ in (\ref{60}) and (\ref{62}) in order to see how the equality works in a higher order  and to check that  there are no hidden underwater stones in this easy proof.

The double pole in eq.(\ref{60}) can be represented as follows
\be
&&\epsilon^{-2}\, (L_2+V_2)=\epsilon^{-2}\sum_{l=2}G^{l-1}_N(L_{2,l}+V_{2,l})=\epsilon^{-2}\sum_{l=2}G^{l-1}_N\sum_{p=1}^{l-1}\frac{p}{l}\, L'_{1,l-p}\cdot h_{1,p}\nonumber \\
&&=\epsilon^{-2}\sum_{l=2}G^{l-1}_N\sum_{p=1}^{l-1}\frac{p}{l}\left((L''_0\times h_{1,p})\cdot h_{1,l-p}+L'_0\cdot(h'_{1,l-p}\times h_{1,p})\right)\, ,
\lb{65}
\ee
where we used the relation (\ref{33}) for $k=1$ and the relation $\beta_{p}=ph_{1,p}\, $ (\ref{8}). In the second line we used the relation obtained by  differentiating w.r.t. metric the equation (\ref{64}):
$L'_{1,l-p}\cdot h_{1,p}=(L''_0\times h_{1,p})\cdot h_{1,l-p}+L'_0\cdot (h'_{1, l-p}\times h_{1,p})$.

On the other hand, in equation (\ref{62}) the double pole is
\be
\epsilon^{-2}\sum_{l=2}G^{l-1}_N\frac{1}{2}\sum_{p=1}^{l-1}\left((L''_0\times h_{1,l-p})\cdot h_{1,p}+2\, \frac{p}{l}\, L'_0\cdot (h'_{1,l-p}\times h_{1,p})\right)\, ,
\lb{66}
\ee
where  we used the second relation in (\ref{8}) to express $h_{2,l}$ in terms of $h_{1,p}\, , \  p=1,\dots, l-1$.

Comparing the expressions (\ref{65}) and (\ref{66}) we  see that the second term, due to $L_0'$, is the same  in both expressions. 
Then we note that
$L''_0$ is symmetric  so that 
\be
(L_0''\times h_{1,p})\cdot h_{1,l-p}=(L''_0\times h_{1,l-p})\cdot h_{1,p}\, .
\lb{67}
\ee
This can be demonstrated by a direct computation of $L''_0$.  Indeed, for two symmetric tensors $A_{ij}$ and $B_{ij}$ one finds that
\be
&&(L''_0\times A)\cdot B=\frac{1}{2}\int (\nabla^a\nabla_i A_{ja}+\nabla^a\nabla_j A_{ia}-\Box A_{ij}-\nabla_i\nabla_j A\nonumber \\
&&-g_{ij}(\nabla^a\nabla^b A_{ab}-\Box A-A_{ab}R^{ab}+\frac{1}{2}AR)+ AR_{ij}-RA_{ij})B^{ij}\nonumber \\
&&=(L''_0\times B)\cdot A\, ,
\lb{67-1}
\ee 
where $A=g^{ij}A_{ij}$.
Eq.(\ref{67})  can be viewed as a symmetric  scalar product   $h_{1,l-p}\star h_{1,p}$.
For a symmetric product like this  one has the
identity
\be
\sum_{p=1}^{l-1}\frac{p}{l}\, h_{1,l-p}\star h_{1,p}=\frac{1}{2}\sum_{p=1}^{l-1} h_{1,l-p}\star h_{1,p}
\lb{68}
\ee
that is the final ingredient needed for the demonstration of  the equality of (\ref{65}) and (\ref{66}).

Some remarks are in order. The relation (\ref{62-0}) may have been  anticipated. Indeed, many authors   have 
noticed  earlier that any counter-term of the form $\int G_{ij}X^{ij}$ that contains at least one power of the Einstein tensor (and, thus, vanishing on-shell) may be absorbed in the original General Relativity  action
by means of a redefinition of the metric, $g_{ij}\rightarrow g_{ij}+G_N X_{ij}$ (see, for instance,  \cite{Metsaev:1986yb} for a relevant discussion). Since the single pole terms $L_{1,l}$ are of this type it is of course natural to expect that  by a similar redefinition all of them (and the higher poles related to $L_{1,l}$ by the RG equations) can be consistently
hidden inside the classical action $L_0$. The metric renormalization  (\ref{2}) could be viewed as a consistent way of doing a redefinition of this type. However, the point in this section is still non-trivial.  The related higher poles  contain
$V_{k,l}\, , \  k\geq 2$ that do not vanish on-shell. Nevertheless, all such terms will go away as soon as  metric in the classical action is redefined as  (\ref{2}), (\ref{8}). The trick is done by the higher order variations of the
classical action $L_0$ that produce terms that are non-vanishing on-shell. That the entire procedure is self-consistent is guaranteed by the renormalization group equations.

\subsection{Renormalization of the total  gravitational action}

A natural question arises whether one can generalize this property to the complete set of counter-terms, i.e. for a generic case of non-vanishing $\{\lambda_l\}$ and $\{v_{1,l}\}$? 
A reasonable guess is that the respective generalization of the classical action should include the higher order terms that are due to the Riemann tensor only,
\be
L_{\rm gr}(g,\lambda)=G_N^{-1}L_0+\sum_{l=2}G^{l-1}_N\, \lambda_l\,P_l\, ,
\lb{69}
\ee
where, additionally to the renormalization of the metric, one has to include the renormalization of the higher order couplings $\{\lambda_l\}$. Notice that the higher order curvature terms include only the terms with the Riemann tensor.
We are now going to show   that this is indeed the right form of the bare gravitational action.   

Consider (\ref{69}) as a function of the bare metric $g_{B}$ and 
the bare coupling constants $\{\lambda^B_l\}$ (\ref{9-2})  and expand in a formal power series as above. It is sufficient to look at the  UV divergent terms  in a single pole (the higher poles are derived from a single pole by the RG equations),
\be
&&L_{\rm gr}(g_B,\lambda_B)=L_{gr}(g_{R},\lambda_R)+\epsilon^{-1}Q_1+\dots\, , \  Q_1=\sum_{l=1}G^{l-1}_N\, Q_{1,l}\, , \nonumber \\
&&Q_{1,1}=L'_0\cdot h_{1,1}\, ,  \ Q_{1,l}=L'_0\cdot h_{1,l}+a_{1,l}P_l+\sum_{p=2}^{l-1}\lambda_p P'_p\cdot h_{1,l-p}\, , l\geq 2
\lb{70}
\ee
The last term in $Q_{1,l}$ is non-trivial for $l\geq 3$.
We now have to incorporate the dependence of all quantities such as  the metric, the beta functions and the terms in the quantum action on the couplings $\{\lambda_p\}$.
Therefore, all these quantities are assumed to be decomposed in the power series with respect to $\lambda$. For a quantity $A_l$ that appears in a loop order $l\geq 2$ one thus has that
\be
A_l=A_l^{(0)}+\sum_{\sm{p=2}}^{\sm{l-1}}A_l^{(p)}\lambda_p+\sum_{\sm{ p_1,p_2=2}}^{\sm{ p_1+p_2=l-1}}A^{(p_1p_2)}_l\lambda_{p_1}\lambda_{p_2}+\dots +\sum_{\sm{ p_1,\dots, p_n=2}}^{\sm{ p_1+\dots+p_n=l-1}}A_l^{(p_1,\dots,p_n)}\lambda_{p_1}\dots\lambda_{p_n}+..
\lb{71}
\ee
In each sum the condition (\ref{z}) is assumed to be satisfied. For any given $l$ there is a finite number of terms in (\ref{71}).
With these definitions we find for $Q_{1,l}\, , \ l\geq 2$
\be
&&Q_{1,l}^{(0)}=L'_0\cdot h^{(0)}_{1,l}+a_{1,l}^{(0)}P_l\, , \nonumber \\
&&Q_{1,l}^{(p)}=L'_0\cdot h_{1,l}^{(p)}+a_{1,l}^{(p)}+P'_p\cdot h^{(0)}_{1,l-p}\, , \nonumber \\
&&Q_{1,l}^{(p_1\dots p_n)}=L'_0\cdot h_{1,l}^{(p_1\dots p_n)}+P'_{p_1}\cdot h^{(p_2\dots p_n)}_{1,l-p_1}+a_{1,l}^{(p_1\dots p_n)}P_l\, , \ n\geq 2
\lb{72}
\ee
Symmetrization over indices $p_1$, $p_2\, , \ \dots$, $p_n$ is assumed in the third line of this equation.
Assuming the form (\ref{71}) for the quantities that enter the beta function equations (\ref{10-1}) and (\ref{10-3}) one finds $\beta_1=h_{1,1}$ for $l=1$ and
\be
&&\hat{\beta}_l^{(0)}=la^{(0)}_{1,l}\, , \  \hat{\beta}_{l}^{(p)}=(l-p)a^{(p)}_{1,l}\, , \ \hat{\beta}_l^{(p_1\dots p_n)}=(l-p_1-\dots -p_n)a^{(p_1\dots p_n)}_{1,l}\, , \ n\geq 2\nonumber \\
&&\beta_l^{(0)}=l h^{(0)}_{1,l}\, , \ \beta^{(p)}_{l}=(l-p)h^{(p)}_{1,l}\, , \ \beta_l^{(p_1\dots p_n)}=(l-p_1-\dots -p_n)h^{(p_1\dots p_n)}_{1,l}\, , \ n\geq 2
\lb{73}
\ee
 for $l\geq 2$.  Notice that if $p_1+\dots+ p_n=l$ then the corresponding beta function vanishes and $a^{(p_1\dots p_n)}_{1,l}$ and $h^{(p_1\dots p_n)}_{1,l}$ are not determined by the equations and can be
 arbitrary. We assume that this ambiguity does not happen. In particular, this is so provided the condition (\ref{z}) holds.
 
The next step is to expand the equation (\ref{16}) in the series (\ref{71}).  First we start with $l=1$ term, see eq.(\ref{17}),
$L_{1,1}-L'_0\cdot \beta_1=0$. Taking into account that $\beta_1=h_{1,1}$ this equation can be re-written as
\be
Q_{1,1}=L_{1,1}\, .
\lb{zzz}
\ee
Then we consider $l\geq 2$.
In zeroth order ($n=0$)  in $\lambda$ one has
\be
lL^{(0)}_{1,l}-L'_0\cdot \beta^{(0)}_l-P_l(\hat{\beta}^{(0)}_l-lv^{(0)}_{1,l})=0
\lb{74}
\ee
that can be re-written, using (\ref{73}), as 
\be
L'_0\cdot h^{(0)}_{1,l}+a^{(0)}_{1,l}P_l=L^{(0)}_{1,l}+v^{(0)}_{1,l}P_l\, .
\lb{75}
\ee
The left hand side of this equation is $Q^{(0)}_{1,l}$ (\ref{72}) and, thus, one has that
\be
Q^{(0)}_{1,l}=L^{(0)}_{1,l}+V^{(0)}_{1,l}\, .
\lb{76}
\ee

In linear order ($n=1$) the equation (\ref{16}), after one used  (\ref{73}) and dropped the overall factor $(l-p)$, reduces to
\be
L^{(p)}_{1,l}-L'_0\cdot h^{(p)}_{1,l}-a^{(p)}_{1,l}P_l+v^{(p)}_{1,l}P_l-P'_p\cdot h^{(0)}_{1,l-p}=0\, .
\lb{77}
\ee
It can be re-written in terms of $Q^{(p)}_{1,l}$ (\ref{72}),
\be
Q^{(p)}_{1,l}=L_{1,l}^{(p)}+V^{(p)}_{1,l}\, ,
\lb{78}
\ee
where $V^{(p)}_{1,l}=v^{(p)}_{1,l}P_l$.

The same algorithm works in $n$-th order of equation (\ref{16}). Using equations (\ref{73}) and dropping the overall factor $(l-p_1-\dots-p_n)$ (that is non-zero due to condition (\ref{z})), one finds
\be
L^{(p_1\dots p_n)}_{1,l}-L'_0\cdot h^{(p_1\dots p_n)}_{1,l}-a^{(p_1\dots p_n)}_{1,l}P_l+ v^{(p_1\dots p_n)}_{1,l}P_l-P'_{p_1}\cdot h^{(p_2\dots p_n)}_{1,l-p_1}=0\, .
\lb{79}
\ee
This equation can be re-written as
\be
Q^{(p_1\dots p_n)}_{1,l}=L_{1,l}^{(p_1\dots p_n)}+V^{(p_1\dots p_n)}_{1,l}\, , n\geq 2
\lb{80}
\ee
where $V^{(p_1\dots p_n)}_{1,l}=v^{(p_1\dots p_n)}_{1,l}P_l$.

Collecting now equations (\ref{zzz}), (\ref{76}), (\ref{78}) and (\ref{80}) one finds that
\be
Q_{1,l}=L_{1,l}+V_{1,l}\, , \ l\geq 1
\lb{81}
\ee
So that a single pole $Q_1$ in the power series expansion of the  bare gravitational action (\ref{70}) is indeed identical to a single pole $L_1+V_1$ in the quantum effective action. This proves our final statement.

\medskip

\noindent{\it \underline{Statement 5.}  The complete set of the UV divergent terms in quantum gravity can be consistently hidden in the bare gravitational action (\ref{69}), that includes  terms of a higher order in the Riemann tensor,  expressed in terms of the bare metric and the bare
higher curvature couplings.}

\bigskip

\noindent This completes the present analysis.

\section{Conclusion}

We have formulated a renormalization group approach to the perturbative quantum gravity based on   't Hooft's method developed earlier for the renormalizable theories.
Our formulation includes the renormalization of the metric, of the higher Riemann  curvature couplings  and the renormalization of the quantum action.
The equations (\ref{10-1}), (\ref{10-3}) and (\ref{33}) form the complete set of  the renormalization group recurrence equations  that can be solved to determine the higher pole counter-terms
in the quantum action. The metric and the higher coupling beta functions are determined by solving eq.(\ref{16}). These equations and, based on them, Statements 1-5 constitute the main result of this paper.
The analysis in the present paper has been  done in the spacetime dimension $d=4$. It is of interest to generalize it to other values of $d$.

We suspect that the approach developed in the present paper  can be extended to other, conventionally considered as non-renormalizable, theories such as a scalar field theory with a non-renormalizable potential
and the interacting theories of the Horndeski type.  We plan to consider these theories elsewhere. 

Our approach may have many applications in quantum gravity. The recurrence equations that we derived  can be used  as a consistency check in a higher loop calculation,
that are  in the case of quantum gravity are very  laborious  and time consuming,  provided  one will be performed in the future. The other possible application is related to the 
computation of the black hole entropy in the perturbative quantum gravity, along the lines developed in \cite{Solodukhin:2019xwx}. 
Finally, it would be interesting to analyze whether one can reconcile the approach developed in this paper with the renormalization ideas suggested in
\cite{Solodukhin:2015ypa}. A related  direction is to extend the present approach to the case of gravity with a non-zero cosmological constant.
This will be considered in a subsequent work.

\section*{Acknowledgements}  I would like to thank M. Shaposhnikov for warm hospitality in his group at EPFL and many inspiring discussions during
the initial stages of this project. I  also acknowledge the useful discussions and communications with D. Kazakov, M. Kalmykov, K. Krasnov, N. Mohammedi, S. Nicolis, S. Sibiriakov and A. Tseytlin.
I thank Amin Faraji Astaneh for his help with  drawing  Figure 1.
The present work is supported in part by the ERC-AdG-2015 grant 694896.

\end{document}